\documentclass[aps,prb, twocolumn,superscriptaddress,amsmath,showpacs,showkeys]{revtex4}
\usepackage{txfonts}
\usepackage[latin1]{inputenc}
\usepackage{graphicx}
\usepackage{dcolumn}
\usepackage{bm}
\usepackage{setspace}

\newcommand {\dg} {\ensuremath{^{\circ}}}

\newcommand {\kvec} {\ensuremath{\mathbf{k}} vector}
\newcommand {\Tz} {\ensuremath{T_{0}}}
\newcommand {\CeRuAl} {CeRu$_{2}$Al$_{10}$}
\newcommand {\CeOsAl} {CeOs$_{2}$Al$_{10}$}
\newcommand {\mSr} {$\mu^+$SR}

\begin{document}

\title{Neutron scattering study of the long-range ordered state in \CeRuAl.}

\author{
Jean-Michel MIGNOT\thanks{E-mail address: jean-michel.mignot@cea.fr}, Julien ROBERT, Gilles ANDR\'{E}, Alexandre M. BATAILLE, Takashi NISHIOKA$^{1}$, Riki KOBAYASHI$^{1}$, Masahiro MATSUMURA$^{1}$, Hiroshi TANIDA$^{2}$, Daiki TANAKA$^{2}$ and Masafumi SERA$^{2}$ \\
}

\affiliation{
Laboratoire L\'{e}on Brillouin, CEA-CNRS, CEA/Saclay, 91191 Gif sur Yvette (France) \\
$^{1}$Graduate School of Integrated Arts and Science, Kochi University, Kochi 780-8520 (Japan) \\
$^{2}$Department of Quantum Matter, ADSM, Hiroshima University, Higashi-Hiroshima, 739-8530 (Japan) \\
}

%

\begin{abstract}
Elastic and inelastic neutron scattering measurements have been performed on powder and single-crystal samples of orthorhombic \CeRuAl. The order forming below $T_0 = 27$ K was identified as a long-range antiferromagnetic state with the wave vector $\mathbf{k} = (1,0,0)$. The magnetic spectral response in the ordered phase, measured on powder, is characterized by a spin gap and a pronounced peak at 8 meV, whose $Q$ dependence suggests a magnetic origin. Both features are suppressed when temperature is raised to \Tz, and a conventional relaxational behavior is observed at 40 K. This peculiar spin dynamics is discussed in connection with recent magnetization results for the same compound.
\end{abstract}


\maketitle


Cerium compounds 	of the Ce\textit{M}$_2$Al$_{10}$ family (\textit{M} = Ru, Os, Fe) exhibit intriguing properties, which do not fit into the general scheme of heavy-fermion of mixed-valence physics. In both \CeRuAl\ and \CeOsAl, the transport coefficients below room temperature are indicative of a Kondo insulating behavior, with a thermally activated variation of the electrical resistivity,\cite{Strydom'09,Nishioka'09} but no reliable estimate of the carrier concentration could be derived from Hall effect measurements.\cite{Tanida'10.2} In this regime, the magnetic susceptibility exhibits a large anisotropy, with the easy direction parallel to the orthorhombic $a$ axis. A phase transition takes place at $T_0 = 27$ and 29 K in \CeRuAl\ and \CeOsAl, respectively, below which the behavior of the material changes dramatically. In \CeRuAl, the resistivity shows a positive jump just below \Tz, then goes through a maximum and drops to a constant, metalliclike---but fairly high residual value below 10 K.\cite{Strydom'09,Nishioka'09} 
There is ongoing controversy over the mechanism of this transition and its order parameter. The initial assumption\cite{Strydom'09} that some type of magnetic order took place, accompanied by the opening of a gap at the Fermi surface, was backed up by the observation of a (quite weak) internal field in muon-spin-resonance (\mSr) experiments.\cite{Kambe'10} However, this interpretation was challenged by $^{27}$Al nuclear quadrupole resonance (NQR) measurements, which revealed no broadening or splitting of the resonance line indicative of an internal field below the ordering temperature.\cite{Matsumura'09} Alternative scenarios have been proposed, such as a charge-density wave,\cite{Nishioka'09} a spin-density wave,\cite{Iizuka'10a} or a dimerization of the Ce sites producing a singlet state,\cite{Tanida'10.1} but each one entails some particular shortcomings. It is therefore prerequisite to know whether long-range magnetic order does occur and, if so, how the magnetic spectral response of the system is affected. This report presents recent neutron scattering measurements aimed at answering these questions. Part of the results were previously published in ref.~\onlinecite{Robert'10}.

\begin{figure}   
\begin{center}
\includegraphics [width=0.75\columnwidth, angle=0] {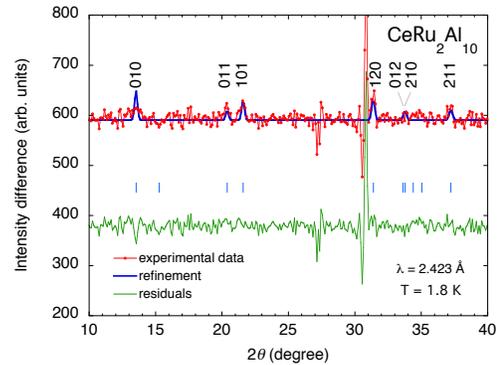}
\end{center}
\caption{(Color online) Refinement of the difference pattern obtained by subtracting the data for $T = 35$ K $> T_0$ from those for  $T = 1.8$ K (ordered state). Experimental data are offset by +600 counts.}
\label{differ_plot}
\end{figure}

Experiment were performed at Orphée-LLB (Saclay) on different instruments: G4-1 diffractometer (800-cell position-sensitive detector) at the incident neutron wavelength $\lambda_i = 2.423$ \AA\ for neutron powder diffraction (NPD), 6T2 four-circle diffractometer at $\lambda_i = 2.35$~\AA\  for single-crystal diffraction (SCD), and 2T thermal-beam triple-axis spectrometer for inelastic neutron scattering (INS). The sample used for the NPD measurements consisted of 14 g of high-quality powder obtained by crushing small single-crystal pieces. It was then combined with an extra 23.7 g of annealed powder, containing a low amount of a second phase, for the INS experiments. A small single crystal grown by the self-flux method, with volume $V \approx 2.7 \times 1.2 \times 1.8$ mm$^3$, was used for the SCD measurements on 6T2.

\begin{figure}   
\begin{center}
\includegraphics [width=0.70\columnwidth, angle=0] {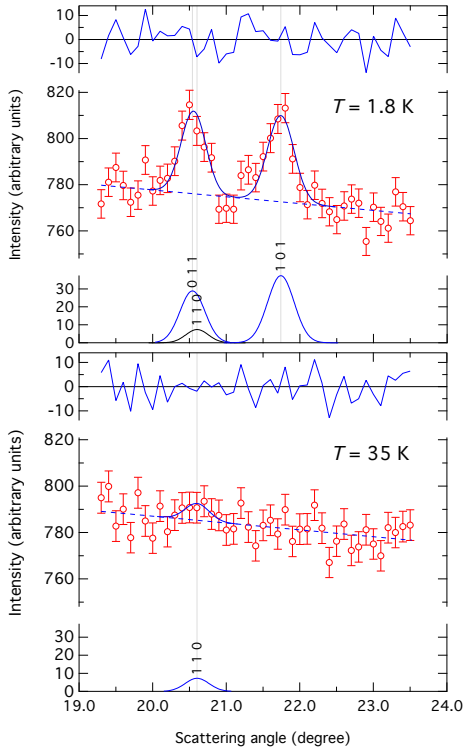}
\end{center}
\caption{(Color online) Low-angle diffraction peaks in \CeRuAl\ measured at $T = 1.8$ (above) and 35 K (below). Each frame shows the experimental data fitted by Gaussian profiles on top of a linear background (middle part), as well as the partial components (bottom part) and the residuals (top part).}
\label{diff_pat}
\end{figure}

\begin{figure}   
\begin{center}
	\includegraphics [width=0.48\columnwidth, angle=0] {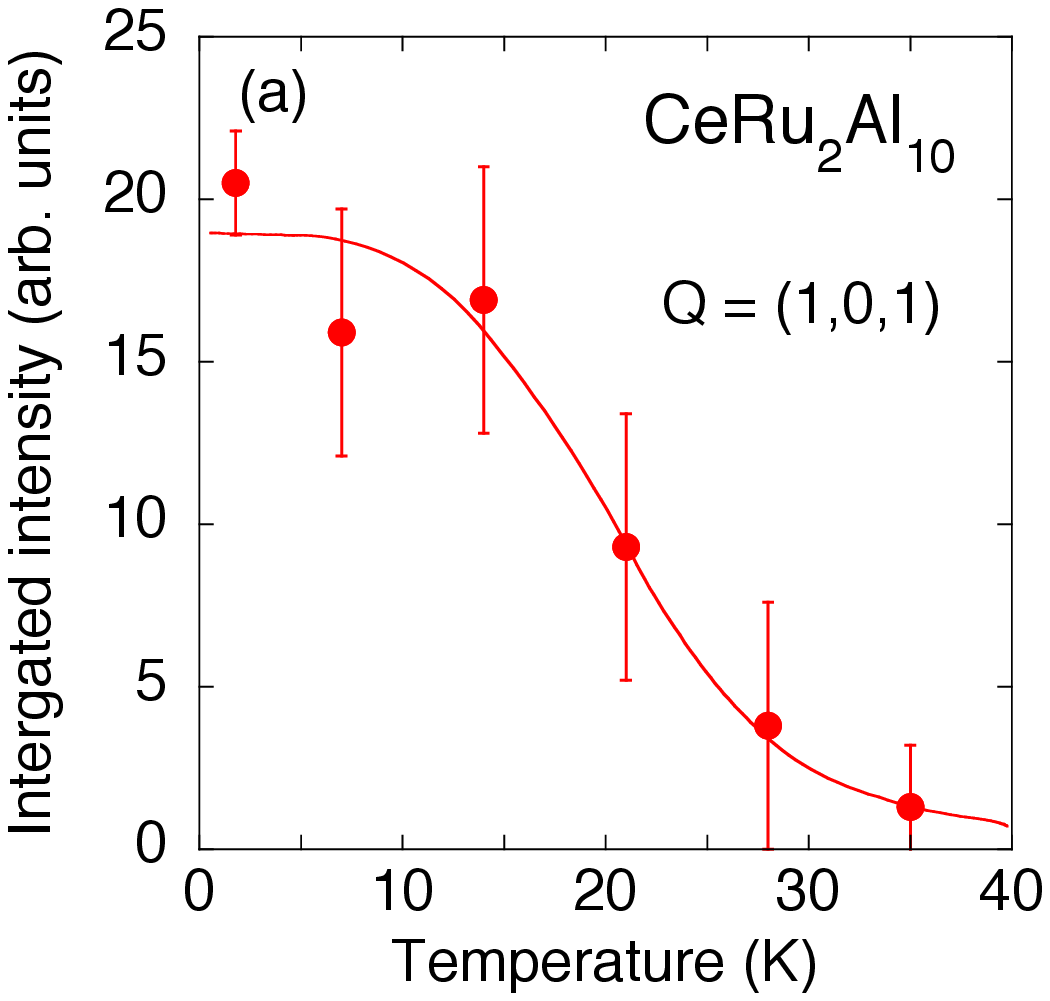}
	\hspace {0.3cm}
	\includegraphics [width=0.42\columnwidth, angle=0] {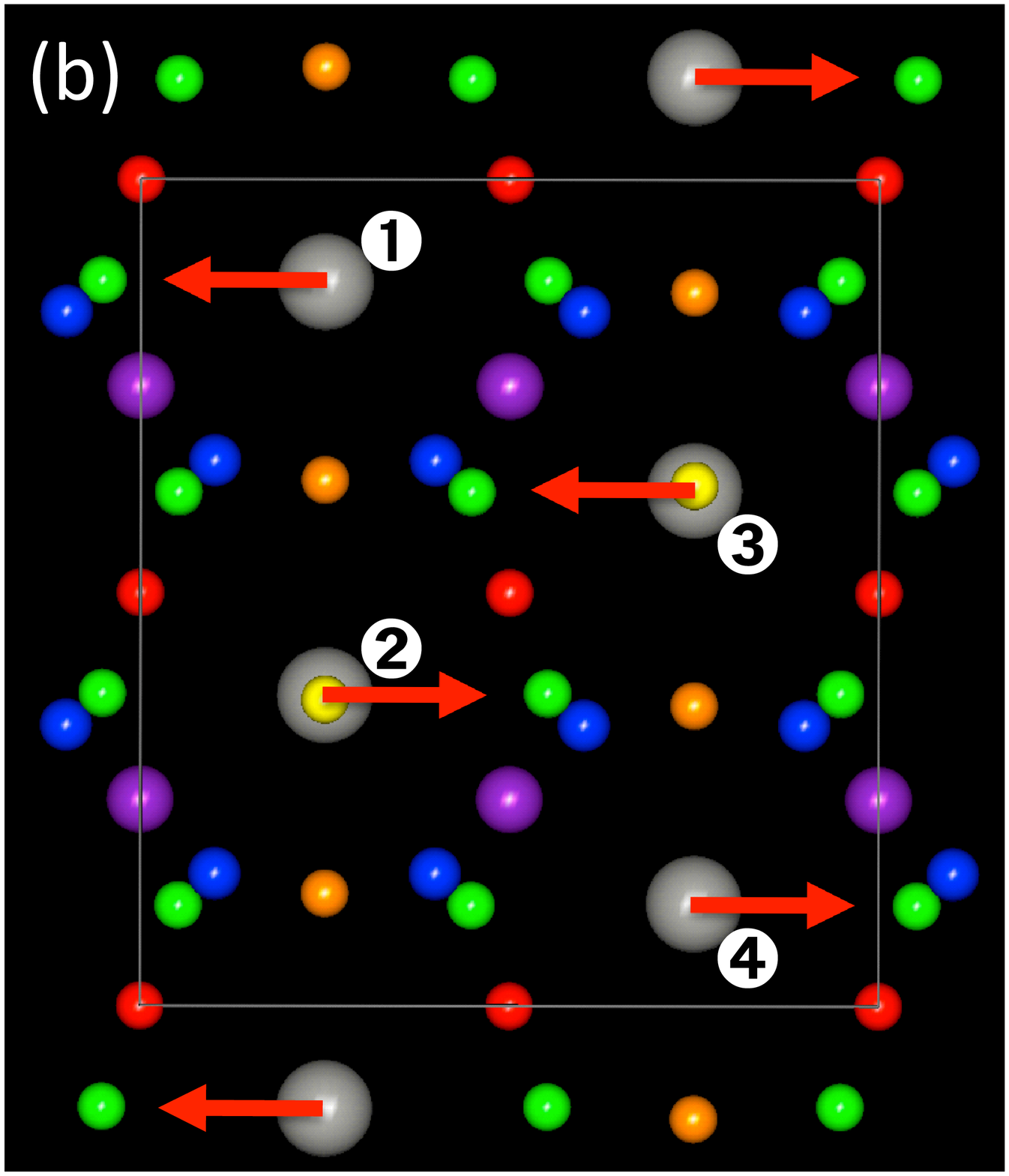}
\end{center}
\caption{(Color online) (a) Temperature dependence of the integrated intensity of the 101 superstructure peak in \CeRuAl\ from the NPD measurement. (b) Projection on the $(b,c)$ plane ($c$ axis horizontal) of the magnetic structure derived from the NPD (\kvec) and SCD (moment orientation) experiments; grey spheres: Ce; purple spheres: Ru; smaller spheres: Al (colors as in ref.~\onlinecite{Tanida'10.2}).}
\label{tdep_101}
\end{figure}

In the NPD pattern measured at $T = 1.8$ K, superstructure peaks are clearly observed at scattering angles corresponding to the 011, 101, and 120 reflections in the original $Cmcm$ crystal structure. On a plot of the difference between the 1.8 K and 35 K data (Fig.~\ref{differ_plot}), the 010 and 211 peaks are further distinguishable. Satellite reflections of higher indices could not be detected within experimental statistics (counting time on the order of 12 or 13 hours), either because they were superimposed on strong nuclear peaks, or because their intensities were too weak. Figure~\ref{diff_pat} shows an expanded view of the 110 (nuclear), 011 and 101 (superstructure) peaks, together with the corresponding data for $T = 35~\mathrm{K} > T_0$. The fit was performed by setting the centers of the peaks to their nominal positions, and fixing the parameters of the weak 110 nuclear reflection to the values obtained from the high-temperature fit. The temperature dependence of the 101 peak intensity plotted in Fig.~\ref{tdep_101}(a) was obtained from NPD patterns recorded at six different temperatures. The signal is seen to vanish close to the transition temperature \Tz, confirming that the superstructure is related to the order parameter. However, the intensities of peaks observed in the present experiment is too low, and their number too small,\cite{hiQ} to allow a reliable comparison with the Ce magnetic form factor, and thus to decide whether they result from magnetic order or a structural distortion. Khalyavin \textit{et al.} have recently reported a time-of-flight NPD study on the same compound,\cite{Khalyavin'10} from which they conclude that the order is indeed magnetic. However, the data presented in their report suffer from similar limitations to ours (few peaks with rather low statistics), and it is definitely worthwhile to seek more stringent evidence, e.g. from polarized neutron measurements.\cite{npol} In the magnetic interpretation, the NPD data imply an antiferromagnetic (AFM) order with a $\mathbf{k}_{\mathrm{AF}} = (1,0,0)$ wavector. A Rietveld refinement\cite{fullprof'01} of the difference pattern over the angular range $3\dg \le 2\theta \le 40\dg$ yields the magnetic structure depicted in Fig.~\ref{tdep_101}(b), with a strongly reduced ordered Ce magnetic moment of about 0.32(4) $\mu_{\mathrm{B}}$ (0.34(2) $\mu_{\mathrm{B}}$ in ref.~\onlinecite{Khalyavin'10}). 

\begin{figure}   
\begin{center}
\includegraphics [width=0.80\columnwidth, angle=0] {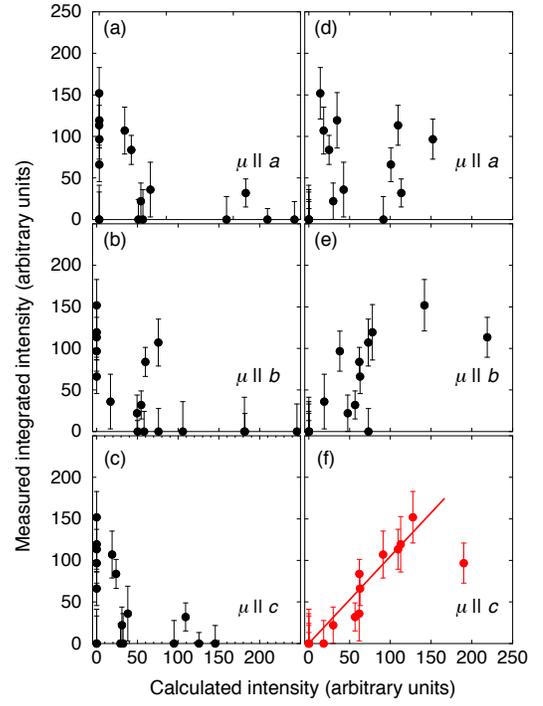}
\end{center}
\caption{(Color online) Comparison of the experimental integrated intensities (corrected for the Lorentz factor) of low-angle superstructure peaks, measured at temperatures between 11 and 18 K, with calculations of the squared magnetic structure factor $|F_{\perp \mathbf{Q}}(\mathbf{Q}|^2$ for different moment directions and relative orientations inside the unit cell: [1, 2, 3, 4] = [+ -- -- +] (left) or [-- + -- +] (right); numbers refer to Ce sites as denoted in Fig.~\ref{tdep_101}(b).}
\label{4circ}
\end{figure}

To ascertain the orientation of the Ce moment, we have performed four-circle experiments on a small single-crystal. Because of temperature control problems and for lack of beam time, data collection in that run was restricted to low scattering angles. Nevertheless, one sees in Fig.~\ref{4circ} that a significantly better agreement between calculated and measured intensities is achieved (frame (f)) when the moments are assumed to be aligned along the $c$ axis, with opposite orientations at the sites (belonging to different AFM sublattices) denoted 1 and 4 in Fig.~\ref{tdep_101}(b). The outlier data point on the right-hand side corresponds to the 120 reflection, whose measured intensity is much weaker than expected from the calculation.

The discrepancy between the easy magnetic ($a$) axis in the paramagnetic state and the orientation of the ordered moments along $c$ revealed by the neutron diffraction measurements is quite striking. Multi-$k$ structures cannot be invoked in the present orthorhombic space group (the ``star'' of the magnetic $k$ vector reduces to $\pm \mathbf{k}_{\mathrm{AF}}$) and one must thus assume some sizable anisotropy to exist in the interactions between Ce ions, though its origin is not presently understood.

\begin{figure}   
\begin{center}
\includegraphics [width=0.90\columnwidth, angle=0] {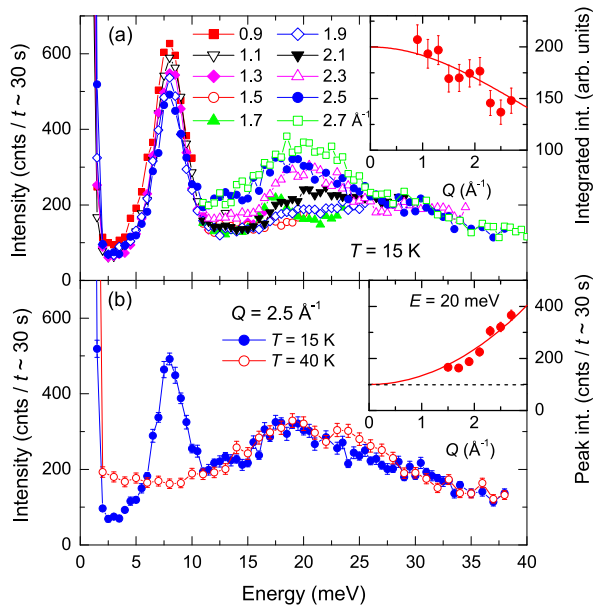}
\end{center}
\caption{(Color online) (a) INS spectra of \CeRuAl\ at $T = 15$ K measured at fixed final energy $E_f =$ 14.7 meV for different momentum transfers (for clarity, some data sets have been restricted to $E > 10$ meV); inset: $Q$ dependence of the integrated intensity of the low-energy peak; solid and dashed lines denote calculations for a single-ion Ce form factor or a Ce-dimer structure factor, respectively. (b) Comparison of spectra below (15 K ) and above (40 K) \Tz; inset: $Q$ dependence of the intensity at the energy of the higher peak, $E = 20$ meV (from ref.~\onlinecite{Robert'10}).}
\label{inelast}
\end{figure}

In order to gain insight into the underlying mechanisms, the magnetic spectral response was studied by inelastic neutron scattering on powder. The results of the thermal-beam measurements, performed at fixed final neutron energy, $E_f = 14.7$ meV, are summarized in Fig.~\ref{inelast}. The main result, shown in frame (a), is the existence of a spin gap in the AFM state at $T_{min} =$ 15 K, with a pronounced peak centered at 8 meV, whose $Q$ dependence is roughly consistent with that expected for the Ce$^{3+}$ single-ion magnetic dipole form factor (upper inset). This peak is broader than the experimental resolution of approximately 1.5 meV at that energy, which may result either from damping, or from powder averaging of a dispersive mode. The spin gap and the excitation at 8 meV both disappear on heating above \Tz, leaving place to a quasielastic-like signal, as can be seen in frame (b). A second, very broad maximum occurs around 20 meV at $T = 15$ K, and it does not change significantly on heating up to 40 K. The increase in its intensity at larger $Q$ values (lower inset) is indicative of one, or several, likely dispersive, lattice excitations.

From its temperature dependence, it is clear that the 8-meV excitation is related to the onset of long-range order below \Tz. Above the transition, the system recovers a relaxational behavior as found in many heavy-fermion or valence-fluctuation compounds. It can be noted that a similar behavior was recently reported for \CeOsAl, with a peak shifted to higher energy (11 meV).\cite{Adroja'10a} In a previous paper, it was argued that the form of the spectral response at 15 K could result from a singlet-triplet transition associated with a Ce dimer, but the probable occurrence of an AFM ordered state in this compound now seems more in favor of a magnon-type excitation with an anisotropy gap. In this interpretation, however, the rather large experimental value of the gap implies a strong anisotropy. As mentioned above, the situation regarding the ordered state is unclear in this respect, since the moments are not oriented along the single-ion easy axis. High-field magnetization experiments\cite{Tanida'10.3} have shown that a spin-flop-like transition occurs around 4 T for $H$ applied along the $c$ axis whereas, for $H \parallel a$, the magnetization increases linearly up to $\sim30$ T. However, a spin-flop threshold of only 4 T seems difficult to reconcile with the existence of an isotropy gap on the order of 8 meV. This apparent contradiction further suggests that the order forming below \Tz\ does not correspond to a conventional AFM state, and emphasizes the need for a detailed study of the $Q$ dependence of the excitation on single crystals when pieces of sufficient size become available.

The authors are grateful to F. Maignen for technical assistance, and to the organizers of ICHE2010 for financial support.

%


\begin{thebibliography}{10}
\bibitem{hiQ}
Reflections at larger Q values are obscured by the overlap with numerous
  nuclear structure peaks.
\bibitem{npol}
Preliminary polarization analysis on the 301 satellite indicates that the
  neutron intensity appears entirely in the spin-flip channel for an incident
  polarization parallel to the scattering vector, and is thus magnetic in
  origin.
\end{thebibliography}

\end{document}